\documentclass[12pt]{iopart}

\usepackage[]{graphics,graphicx,epsfig}

\usepackage{enumerate}
\usepackage{amsfonts}
\usepackage{amssymb}
\usepackage{amsthm}

\begin{document}


\title{Pre- and post-selection paradoxes in quantum walks}

\author{T Kopyciuk, M Lewandowski and P Kurzy{\'n}ski}
\address{Faculty of Physics, Adam Mickiewicz University in Pozna\'n, Umultowska 85, 61-614 Pozna\'n, Poland}
\ead{pawel.kurzynski@amu.edu.pl}


\begin{abstract}
Many features of single-partite quantum walks can be simulated by classical waves. However, it was recently experimentally shown that some temporal sequences of measurements on a quantum walker do not admit a classical description in terms of macro-realistic theories. Here, we extend this observation and present pre- and post-selection scenarios in quantum walks leading to logical paradoxes. Such paradoxes were recently shown to be equivalent to proofs of contextuality, therefore we provide an additional argument for non-classicality of a quantum walk model. The assumptions behind the claims of non-classicality (including contextuality and violations of macro-realism) are also discussed. 
\end{abstract}


\section{Introduction}

In a classical random walk a particle's path is determined by consecutive coin tosses. Such a process generates a diffusive Gaussian spread. A quantum analogue of this process uses a quantum coin and a quantum particle. A unitary coin toss and an outcome dependent shift leave the system in a superposition of moving to the right with the coin showing heads and moving to the left with the coin showing tails. After many unitary coin tosses and shifts a peculiar ballistic spatial distribution is obtained \cite{Aharonov,Review1,Review2,Review3,Review4}.    

Although the above model arose within quantum physics, it soon became clear that many quantum walk features can be simulated with classical waves \cite{ClassicalQW1,ClassicalQW2,ClassicalQW3,ClassicalQW4}. But does it mean that such model does not exhibit any non-classical effects? The answer appears to be negative because in a recent experiment \cite{QWLG} it was shown that single-particle quantum walks can be used to observe a violation of the temporal Bell inequalities of the Leggett-Garg type. This means that, unlike huge classical macroscopic objects, quantum walkers do not admit a macro-realistic description \cite{LG}. 

To further understand classical and quantum features of the quantum walk model one should try to observe within it other effects that are deemed non-classical. Here, we focus on observation of the pre- and post-selection paradoxes \cite{ABL}, that were recently shown to be equivalent to proofs of contextuality \cite{Pusey,PuseyLeifer}. In fact, we show that single-particle quantum walks are natural models to study the pre- and post-selection (PPS) scenarios since due to their discrete-time dynamics PPS paradoxes can be observed at every time step.

The PPS scenarios are based on preparing (pre-selecting) the system in some state $|\psi\rangle$ and later measuring it (post-selecting) in some other state $|\phi\rangle$. Next, one tries to predict outcomes of measurements that could be in principle performed after pre-selection and before post-selection. These counterfactual predictions were shown to lead to bizarre effects, such as the 3-box paradox \cite{3box}. 

Note that PPS scenarios were already discussed in the quantum walk setting. In the seminal work by Aharonov, Davidovich and Zagury \cite{Aharonov} it was shown that a particular preparation and well designed measurement can result in a counterintuitive distortion of the spatial probability distribution after a single step. Here, we go further and propose a collection of logical and dynamical pre- and post-selection paradoxes. We show that such paradoxes can occur at every step of the quantum walk. In addition, we show that PPS quantum walks give rise to a peculiar counterfactual discrete-time dynamics. 

Finally, we need to mention that by linking non-classical quantum walk features with PPS effects we enter a controversial territory \cite{debate1,debate2,debate3,debate4,debate5,debate6}. It may seem that works \cite{Pusey,PuseyLeifer} liberate the PPS effects from at least some controversies by equating it to the well established non-classical feature such as contextuality \cite{KS} (or more precisely, as argued in \cite{PuseyLeifer}, contextuality as defined by Spekkens \cite{SpekkensContext}). Nevertheless, one needs to be aware that the problem of contextuality in quantum physics, as well as the problem of macro-realism \cite{LG} studied in the aforementioned quantum walk experiment \cite{QWLG}, is itself not free from controversies. We are going to discuss this in more details below. 


\section{Quantum walks}

\subsection{Discrete-time quantum walks}

\begin{figure}[t]
\center{\includegraphics[scale=0.6]{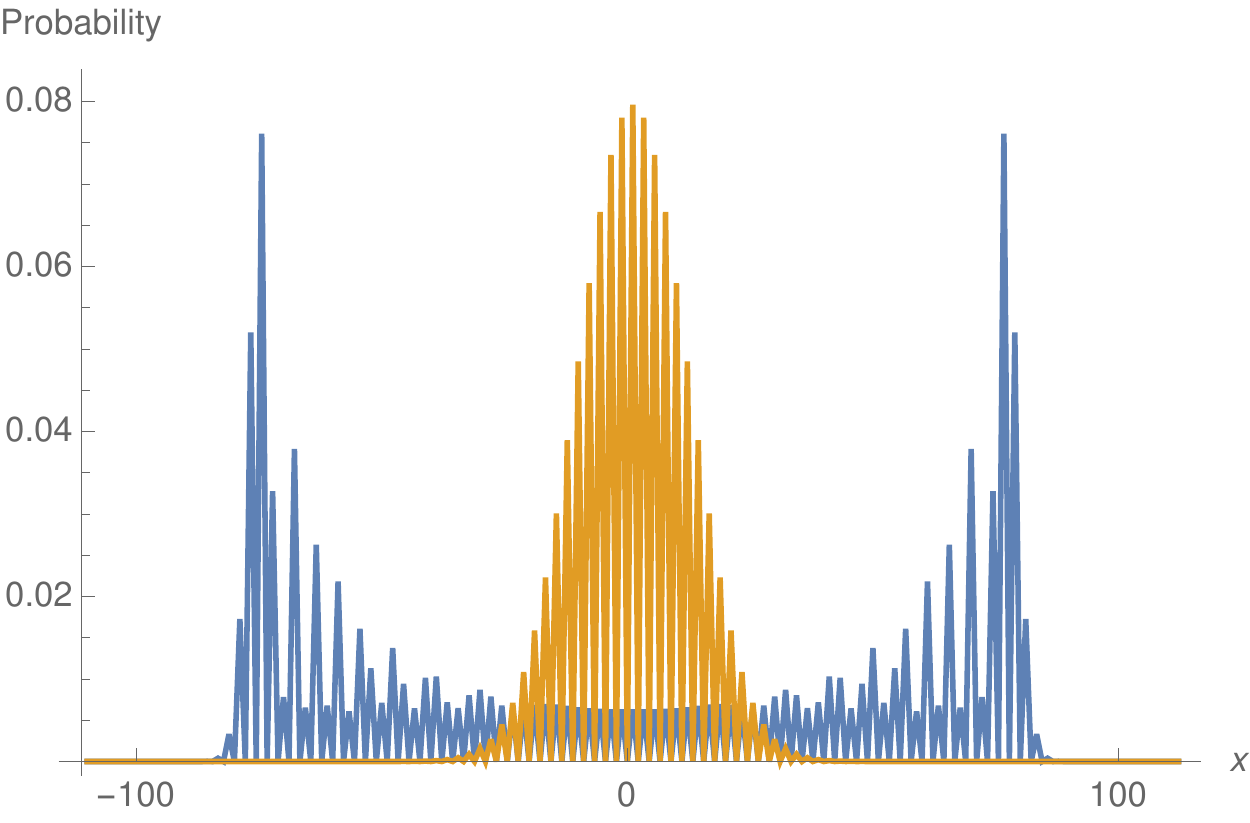}}
\caption{Spatial probability distributions of the Hadamard walk (blue) and the classical random walk (orange) after 100 steps. Both processes start at $x=0$ and the initial coin state of the Hadamard walk is $\frac{1}{\sqrt{2}}(|0\rangle + i|1\rangle)$. \label{fig1}}
\end{figure}

A discrete-time quantum walk (DTQW) in its standard version consists of a particle, whose position is determined by a single integer ($x\in {\mathbb Z}$), and a two-state coin described by a binary variable ($c=0,1$). The state of the system is
\begin{equation}
|\psi\rangle = \sum_{x,c} \alpha_{x,c} |x\rangle \otimes |c\rangle,
\end{equation}
where $\alpha_{x,c}$ are probability amplitudes. Its evolution is governed by a unitary transformation
\begin{equation}
U=S(I_x \otimes C),
\end{equation}
where $I_x$ is the identity operator on the position space, $S$ is a conditional translation operator
\begin{equation}
S|x\rangle\otimes |c\rangle = |x+(-1)^c\rangle \otimes |c\rangle
\end{equation}
and $C$ is a coin 'toss' operator. The operator $C$ is often taken to be the Hadamard transformation $H$
\begin{eqnarray}
H|0\rangle &=& \frac{1}{\sqrt{2}}(|0\rangle + |1\rangle), \\
H|1\rangle &=& \frac{1}{\sqrt{2}}(|0\rangle - |1\rangle),
\end{eqnarray}
hence the corresponding DTQW is dubbed the Hadamard walk. After $T$ steps the initial state $|\psi_0\rangle$ is transformed into
\begin{equation}
|\psi_T\rangle = U^T|\psi_0\rangle.
\end{equation}
The corresponding spatial probability distribution is 
\begin{equation}
p(x,T)=\langle \psi_T | (|x\rangle\langle x|\otimes I_c)|\psi_T\rangle,
\end{equation}
where $I_c$ is the two-dimensional identity operator on the coin space. This distribution differs from the classical Gaussian one (see Fig. \ref{fig1}). Apart from its anti-Gaussian shape, the DTQW distribution is ballistic, i.e., its standard deviation is proportional to $T$. 

\subsection{Multi-coin quantum walks}

Multi-coin quantum walks (MCQWs) were introduced in \cite{MCQW1,MCQW2}. This time the system consists of a particle and more than one coin. Here, we focus on a special case for which the number of coins is equal to the number of steps $T$. The state of the system is of the form
\begin{equation}
|\psi\rangle = \sum_{x,c_1,\ldots,c_T} \alpha_{x,c_1,\ldots,c_T} |x\rangle \otimes \left(\bigotimes_{j=1}^T |c_j\rangle\right),
\end{equation}
where each $c_j=0,1$. The $i$-th step of the evolution is governed by a unitary transformation
\begin{equation}\label{Ui}
U_i=S_i(I_x\otimes I_c \otimes \ldots \otimes C_i \otimes \ldots I_c),
\end{equation}
where $S_i$ is the position shift conditioned on the $i$-th coin
\begin{equation}
S_i |x\rangle \otimes \left(\bigotimes_{j=1}^T |c_j\rangle\right) = |x+(-1)^{c_i}\rangle \otimes \left(\bigotimes_{j=1}^T |c_j\rangle\right)
\end{equation}
and $C_i$ is the coin toss of the $i$-th coin. After $T$ steps we get
\begin{equation}
|\psi_T\rangle = \prod_{i=1}^T U_{i} |\psi_0\rangle.
\end{equation}  

Due to commutation relations between conditional shifts and coin toss operations all of the latter can be applied before the particle starts to move. Note, that for $i\neq j$ the following holds 
\begin{eqnarray}
& &S_i(I_x\otimes I_c \otimes \ldots \otimes C_j \otimes \ldots I_c)= \nonumber \\
& &(I_x\otimes I_c \otimes \ldots \otimes C_j \otimes \ldots I_c)S_i.
\end{eqnarray}
This allows us to write
\begin{equation}\label{MCQWevolution}
|\psi_T\rangle = \left( \prod_{i=1}^T S_i \right) \left(I_x\otimes \left(\bigotimes_{i=1}^T C_i \right)\right)|\psi_0\rangle.
\end{equation}
In fact, coin toss operations can be included in the preparation of the initial state
\begin{equation}
|\phi_0\rangle \equiv \left(I_x\otimes \left(\bigotimes_{i=1}^T C_i \right)\right)|\psi_0\rangle,
\end{equation}
which gives
\begin{equation}\label{coinless}
|\phi_T\rangle \equiv |\psi_T\rangle = \left( \prod_{i=1}^T S_i \right)|\phi_0\rangle.
\end{equation}


\subsection{Walking automaton}

\begin{figure}[t]
\center{\includegraphics[scale=0.25,trim={-3cm 3cm 3cm 1cm},clip]{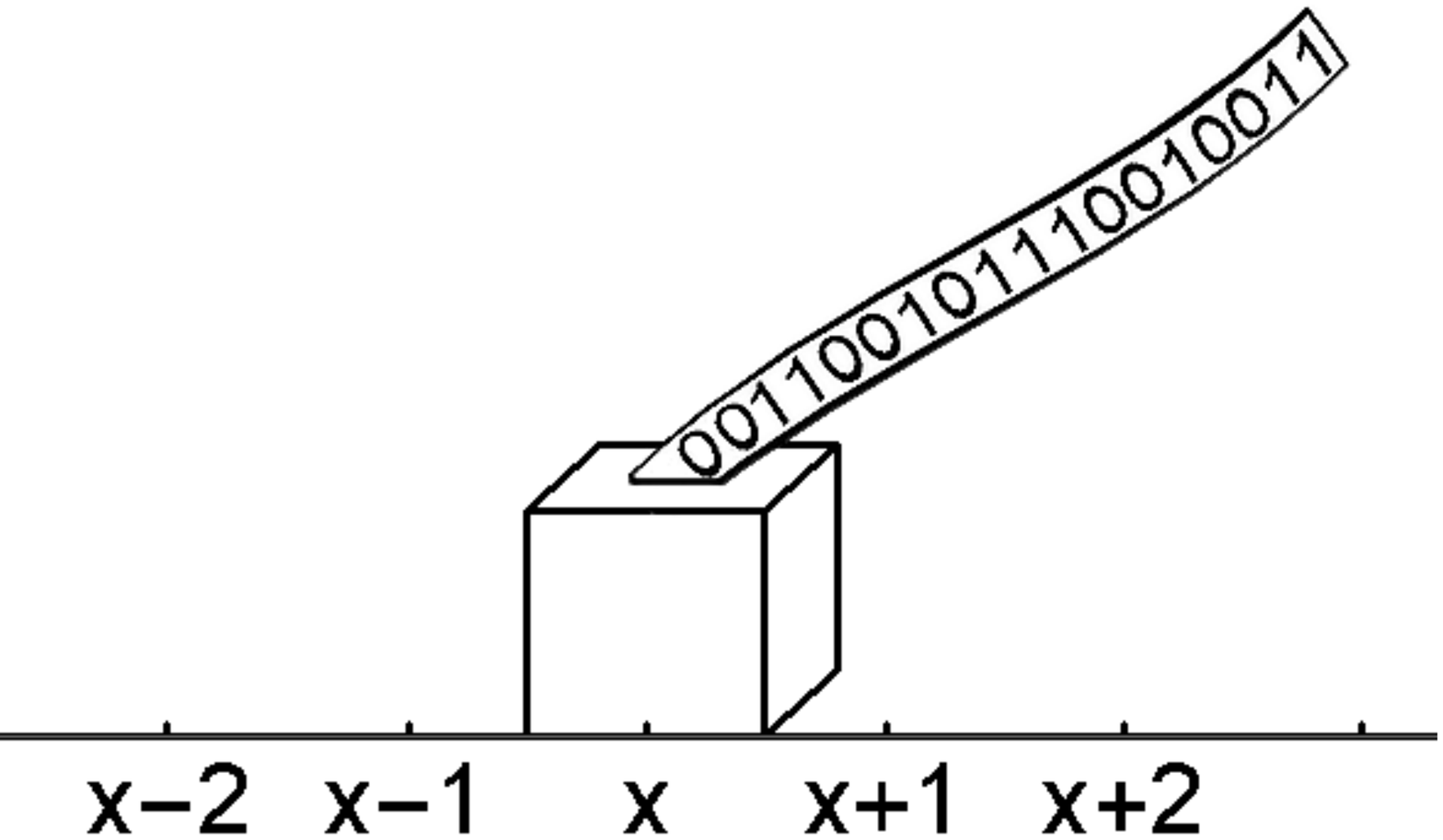}\includegraphics[scale=0.25,trim={1cm 1cm 1cm 1cm},clip]{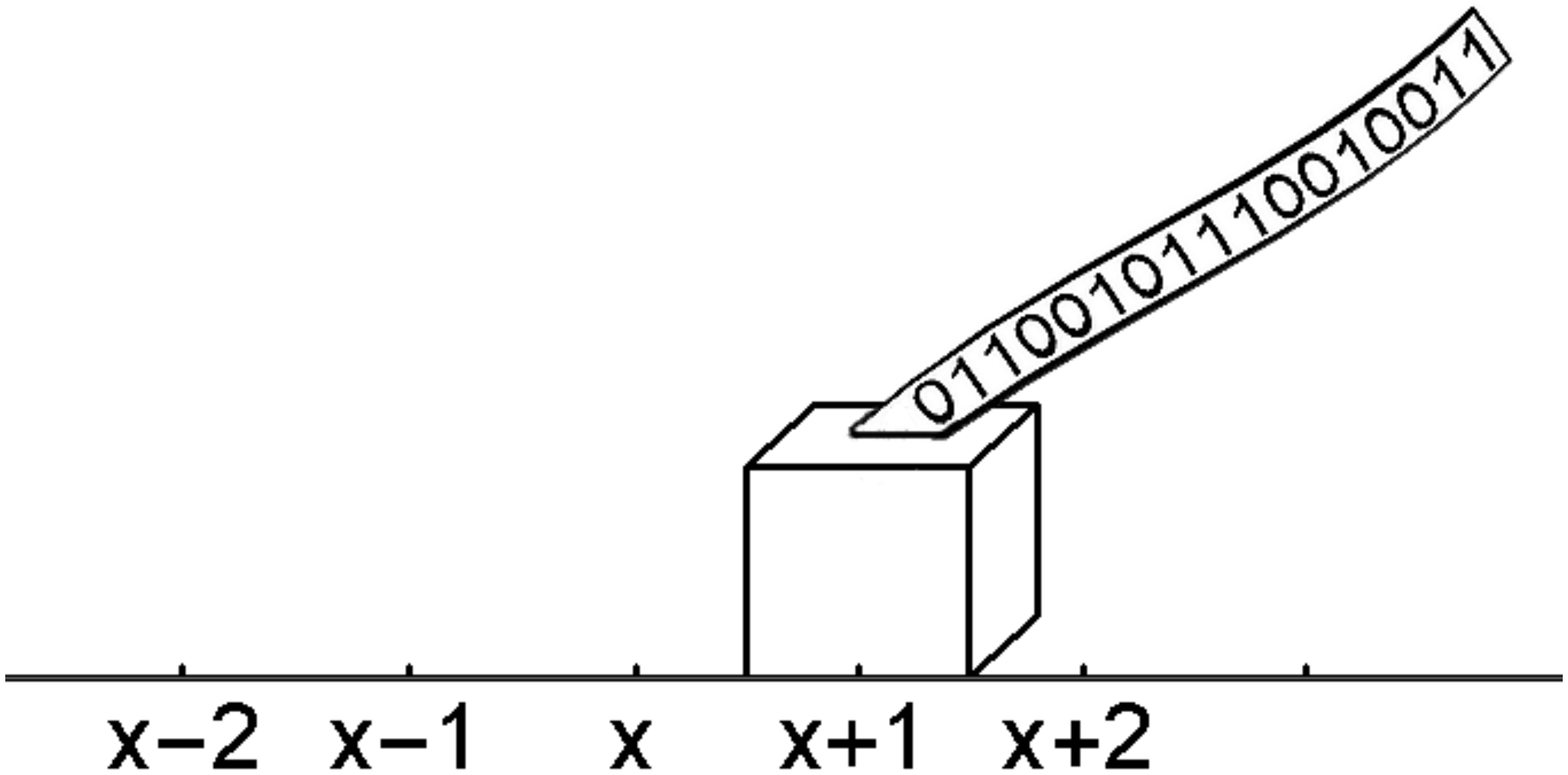}
\includegraphics[scale=0.25,trim={-4cm 3cm 0cm 1cm},clip]{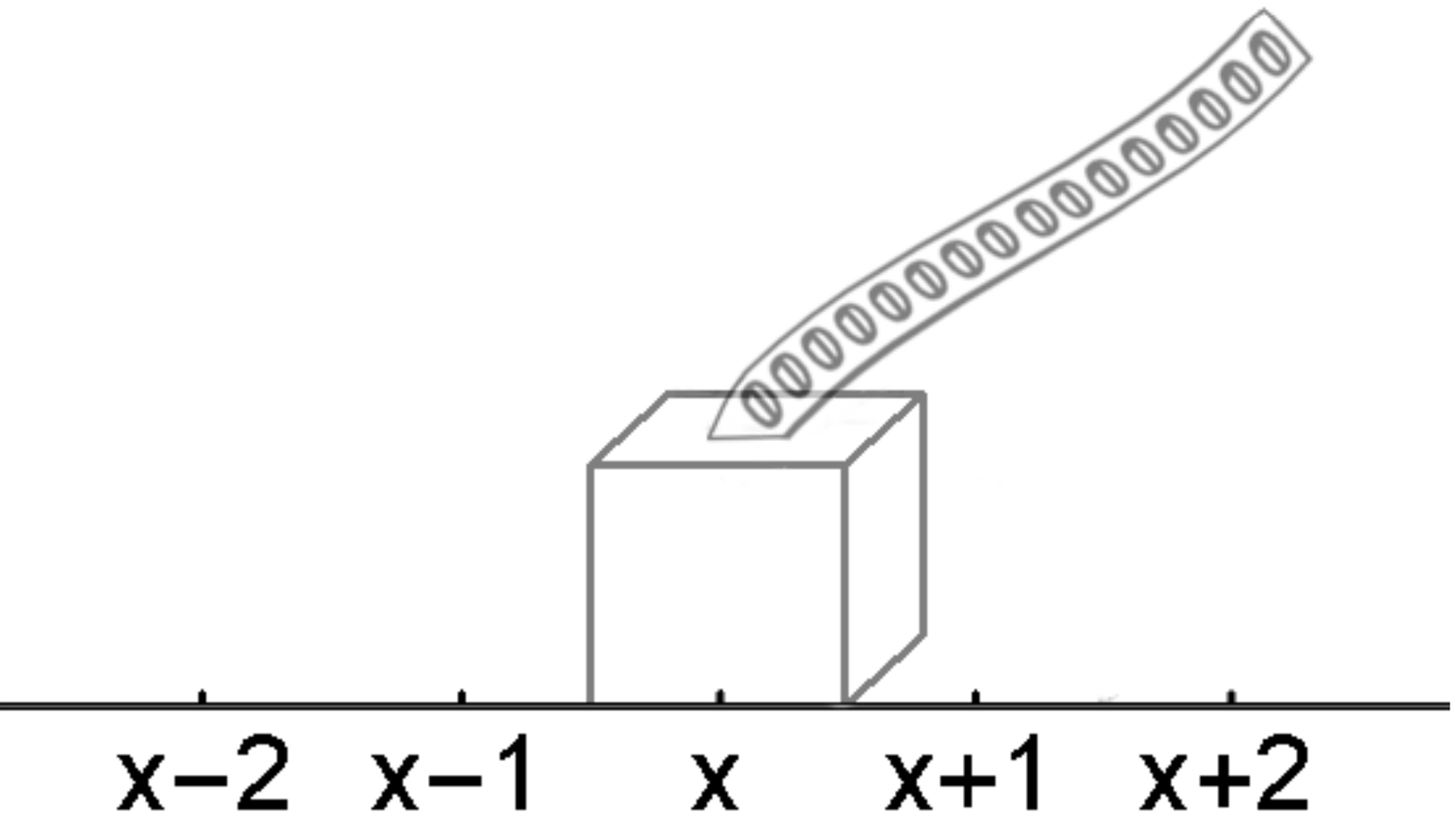}\includegraphics[scale=0.25,trim={1cm 3cm 1cm 1cm},clip]{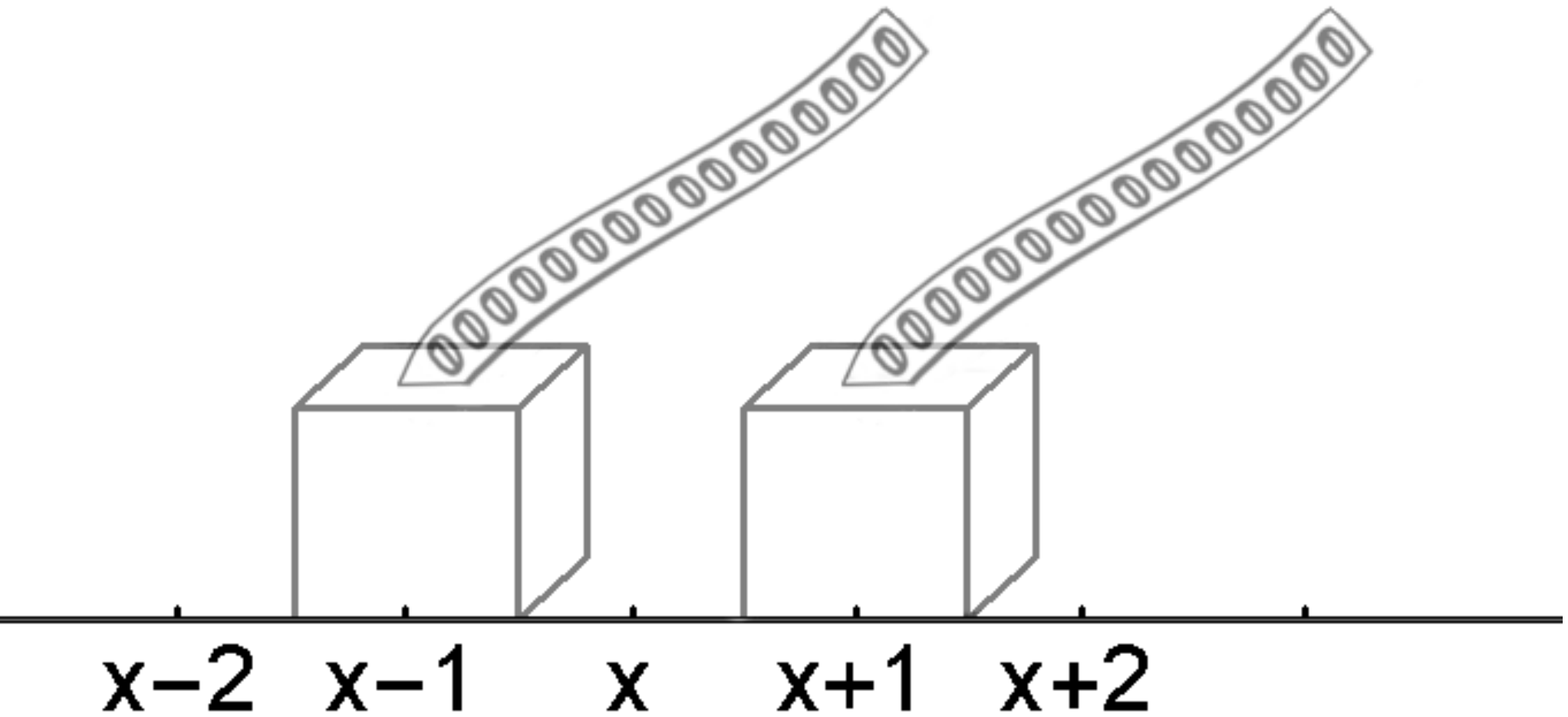}}
\vspace{-5mm}
\caption{MCQW as a walking automaton. A string of bits or qubits is fed into it and the automaton moves left or right. For example, the automaton at position $x$ (top left) is fed zero and moves one step to the right (top right). In case of quantum inputs (bottom left) the automaton can go into a superposition of moving right and left (bottom right). \label{fig2}}
\end{figure}

From now on we skip coin toss operations and use the initial state $|\phi_0\rangle$. The evolution given by (\ref{coinless}) allows us to interpret MCQW as an automaton capable of moving left or right depending on the binary input that is fed into it (see Fig. \ref{fig2}). A bit string encodes a path the automaton is going to follow. For example, the string $00101\ldots$ fed into the automaton will cause it to move: right, right, left, right, left, $\ldots$ If instead of a bit string one prepared a string of qubits (encoded on the initial multi-coin state) then the automaton would go into superposition and would coherently walk along many paths. Note that the string of qubits could be prepared in a multi-partite entangled state. Such a preparation would lead to visible differences in the resulting spatial distribution \cite{MCQWent}. 

Finally, note that if the scenario considers automaton moving all the time left, all the time right, or a superposition of these two possibilities, the MCQW can be replaced by the DTQW in which the coin toss operation $C=I_c$. In this case the path is encoded on a single qubit that is repeatedly fed into the automaton. 


\section{Paradoxes and controversies}

\subsection{Two-state vector formalism}

Quantum dynamics of a closed system is usually formulated in the following way. At time $t=0$ the system is initialized in a state $|\psi_0\rangle$, then it evolves for time $T$, where the evolution is governed by a unitary operator $U(T)|\psi_0\rangle = |\psi_T\rangle$, and finally it is measured. If the measurement is represented by a set of projectors $\{|\chi_i\rangle\langle\chi_i|\}$, then after the measurement the state of the system is $|\chi_i\rangle$ with probability $p_{\chi_i}(T)=|\langle \chi_i|\psi_T\rangle|^2$. Note that apart from probabilistic measurement process the evolution of the system is deterministic and time symmetric, therefore it can be reversed $U^{\dagger}(T)|\psi_T\rangle=|\psi_0\rangle$. It is commonly accepted that the complete description of the system at time $0<t<T$ is given by $|\psi_t\rangle = U(t)|\psi_0\rangle$. Hence, given the unitary evolution operator $U(t)$, prior to the measurement the properties of the system are fully determined by the initial conditions $|\psi_0\rangle$.  

However, it was suggested that $|\psi_0\rangle$ may not be enough to describe the properties of the system prior to the measurement \cite{Watanabe,ABL}. It was proposed that one should also take into account the post-measurement state. Let us assume that the result of the measurement was $|\chi_i\rangle$. In this case one is given two temporal boundary conditions $|\psi_0\rangle \equiv |pre\rangle$, which is called the {\it pre-selected} state, and $|\chi_i\rangle \equiv |post\rangle$, which is called the {\it post-selected} state. Given two empirical facts, namely that the system was prepared in the state $|pre\rangle$ and later measured in the state $|post\rangle$, one infers what were the properties of the system after the preparation and before the measurement. Since the state $|pre\rangle$ can be evolved forward in time $|pre(t)\rangle = U(t)|pre\rangle$ and the state $|post\rangle$ can be evolved backward in time $|post(t)\rangle=U^{\dagger}(T-t)|post\rangle$, the full description of the system at time $t$ is given by both, $|pre(t)\rangle$ and $|post(t)\rangle$. This approach is known as the two-state vector formalism (TSVF) \cite{ABL}.


\subsection{Contextuality in logical pre- and post-selection paradoxes}

The TSVF led to intriguing PPS paradoxes \cite{PPParadoxes}, such as the 3-box paradox \cite{3box}, superluminal velocities \cite{SuperluminalQW}, quantum Cheshire cat \cite{QCC}, quantum pigeonhole effect \cite{QPE}, or disappearing and reappearing particles \cite{DRP}, just to name a few. These paradoxes gave rise to a number of controversies and their true 'paradoxical' nature is still debated \cite{debate1,debate2,debate3,debate4,debate5,debate6}. 

One of such controversies is related to contextuality. Initially the TSVF was thought to impose direct contextuality of quantum theory in a sense that the probabilities of certain events derived for PPS states (see Eq. (\ref{ABLprob}) below) were shown to depend on which other compatible measurements were performed at the same time \cite{AharonovContext}. However, Bub and Brown \cite{BubBrown} rejected such a possibility by showing that a generalization of this scenario to a spatially separated bipartite setting would imply direct nonlocality of quantum theory allowing for superluminal communication. 

Almost two decades later Leifer and Spekkens showed \cite{LeiferSpekkens} that PPS paradoxes are linked to proofs of contextuality of quantum theory, i.e., an inability of a theory to be described by a non-contextual hidden variable model \cite{KS,SpekkensContext}. Interestingly, this inability can occur in theories in which probabilities do not depend on the measurement context, which is a generalization of the no-signalling principle. Quantum mechanics is a flagship example of such a theory. Still, the authors of the above result noted that PPS paradoxes can be simulated by non-contextual theories once measurement disturbance is included \cite{debate3}. 

However, recently Pusey \cite{Pusey} and then Pusey and Leifer \cite{PuseyLeifer} proved that disturbances caused by quantum measurements are not strong enough to be solely responsible for the PPS paradox and thus such paradox is equivalent to a proof of contextuality. In particular, here we focus on the result presented in \cite{PuseyLeifer}, where it was shown that the above is true under three specific conditions. These conditions are: 1) non-orthogonality of the pre-selected state and the post-selected one, 2) intermediate measurements cause wave-function collapse according to the von Neumann-L{\"u}ders rule, 3) the conditional probabilities assigned to the intermediate measurements are either zero or one.  

The name 'logical PPS paradox' originates from condition 3)  \cite{LeiferSpekkens,PuseyLeifer}. In this case the possible intermediate measurements are assigned deterministic conditional outcomes. Imagine that at time $t$ one can choose to measure either an event A or an event B. The corresponding measurements are described by pairs of projectors $\{\Pi_A,I-\Pi_A\}$ and $\{\Pi_B,I-\Pi_B\}$, respectively. Moreover, assume that these two events are exclusive, which implies $\Pi_A \Pi_B = 0$. The conditional probability that event $X$ ($X=A,B$) is registered at time $t$, given PPS states, is \cite{ABL}
\begin{eqnarray}
& &p(\Pi_X|pre(t),post(t)) = \label{ABLprob} \\
& &\frac{|\langle post(t)|\Pi_X|pre(t)\rangle|^2}{|\langle post(t)|\Pi_X|pre(t)\rangle|^2+|\langle post(t)|(I-\Pi_X)|pre(t)\rangle|^2}  \nonumber
\end{eqnarray} 
The logical PPS paradox consists of states $\{|pre(t)\rangle,|post(t)\rangle\}$ ($\langle post(t)|pre(t)\rangle\neq 0$) and projectors $\{\Pi_A,\Pi_B\}$ such that probabilities (\ref{ABLprob}) take values either 0 or 1. The paradox occurs when conditional probabilities of both exclusive events are assigned value $1$ 
\begin{equation}
p(\Pi_A|pre(t),post(t))=p(\Pi_B|pre(t),post(t))=1.
\end{equation}
This happens when
\begin{equation}
|\langle post(t)|(I-\Pi_A)|pre(t)\rangle|^2=|\langle post(t)|(I-\Pi_B)|pre(t)\rangle|^2=0.
\end{equation}

Let us recall the renowned 3-box paradox \cite{3box}. In this scenario there is no time evolution between pre- and post-selection, i.e., $U(t)=I$. The system consists of a single particle that can be localized either in box A, or box B, or C and the corresponding quantum states are $\{|A\rangle,|B\rangle,|C\rangle\}$. The system is pre-selected in the state
\begin{equation}
|pre\rangle = \frac{1}{\sqrt{3}}(|A\rangle+|B\rangle+|C\rangle)
\end{equation}
and then it is measured and post-selected in the state
\begin{equation}
|post\rangle = \frac{1}{\sqrt{3}}(|A\rangle+|B\rangle-|C\rangle).
\end{equation}
The probability of post-selection is $|\langle pre| post \rangle|^2=1/9$. 

What if between pre- and post-selection somebody opened box A? According to pre-selection, if the particle was found there, the state of the system would collapse onto $|A\rangle$. If not, the system would collapse onto $|\bar{A}\rangle = \frac{1}{\sqrt{2}}(|B\rangle + |C\rangle)$. This collapse rule is known as the von Neumann-L{\"u}ders rule. However, since it is guaranteed that the system is post-selected in the state $|post\rangle$ and $\langle \bar{A}|post\rangle = 0$ we conclude that the particle would certainly be found in A. This conclusion stems from the conditional probability $p(\Pi_A|pre,post)=1$, where $\Pi_A=|A\rangle\langle A|$. But what if somebody opened box B instead of A? Using similar arguments, if the particle was not found in B the state of the system would collapse onto $|\bar{B}\rangle = \frac{1}{\sqrt{2}}(|A\rangle + |C\rangle)$. However, since $\langle \bar{B}|post\rangle = 0$ we conclude that the particle would be found in B. Again, this stems from the conditional probability $p(\Pi_B|pre,post)=1$, where $\Pi_B=|B\rangle\langle B|$. The paradoxical observation is that a single particle is concluded to be in two different boxes at the same time, which according to \cite{PuseyLeifer} is also a proof of contextuality.

Before we proceed, let us discuss one important issue mentioned earlier. The works \cite{Pusey,PuseyLeifer} show that under certain conditions PPS paradoxes can be equated to proofs of contextuality. This result is probably interpreted by many as an elevation of PPS paradoxes to a level of truly non-classical phenomena. Nevertheless, one needs to be aware that the non-classicality of contextuality \cite{KS}, as well as non-classicality of violations of macro-realism \cite{LG} (observed in the quantum walk experiment \cite{QWLG}), is also a delicate subject. The reason is that both, non-contextuality and macro-realism, rely on an assumption that the measurements only reveal the state of the system. In fact, the same assumption holds for local realism that is tested in Bell scenarios \cite{Bell}. However, unlike in Bell scenarios, it is much harder to support this assumption since different measurements are done on the same system, not on the spatially separated ones, therefore the system can in principle remember which measurement was performed and adapt its responses to future measurements. One can simulate contextuality and violations of macro-realism on classical systems with additional memory. Therefore, local measurements lead to non-classical phenomena under the assumption that the system has no (or limited) memory \cite{Memory1,Memory2}. This is similar to the Bell scenario assumption that there is no superluminal communication between the subsystems, since violation of local realism can be simulated if classical systems are allowed to communicate (locality loophole is not closed). 


\section{Results}

We are going to consider MCQWs and DTQWs initialized in the state $|pre\rangle$, then evolving for $T$ steps and finally measured and post-selected in the state $|post\rangle$. Next, we are going to infer what would happen if some measurement was done after the pre-selection and before the post-selection. Although in principle any measurement can be considered, we limit ourselves to position measurements since they are the most natural ones in the quantum walk model. More precisely, we are going to consider events A, B, $\ldots$ that correspond to finding the particle at position $x_A$, $x_B$, $\ldots$. These events will be represented by projectors $\Pi_x=|x\rangle\langle x|\otimes I_c$, where as before $I_c$ is the identity on the coin/multi-coin register. Below we present four scenarios. 


\subsection{Scenario 1: Long-distance leaps}

We consider an MCQW scenario that partially resembles the 3-box paradox, but is different due to non-trivial time evolution and entanglement between the position and the multi-coin register. Let us pre-select the system in the state
\begin{equation}
|pre\rangle \equiv |pre(0)\rangle = \frac{1}{\sqrt{3}}|x=0\rangle\otimes\left(|A\rangle+|B\rangle+|C\rangle\right),
\end{equation}
where $|A\rangle$, $|B\rangle$ and $|C\rangle$ are multi-coin states that encode three paths. The particle is initially localized at $x=0$ and the multi-coin states, although different, all lead from $x=0$ to $x=s$. The system takes $T$ steps, therefore the encoded paths correspond to bit strings of length $T$, each having $\frac{T+s}{2}$ zeros and $\frac{T-s}{2}$ ones. We fix $s=\frac{T}{3}$, hence $\frac{T+s}{2}=\frac{2T}{3}$ and $\frac{T-s}{2}=\frac{T}{3}$. Moreover, we set 
\begin{eqnarray}
|A\rangle &=& |\underbrace{00\ldots 0}_{\frac{2T}{3}}\underbrace{11\ldots 1}_{\frac{T}{3}}\rangle, \\
|B\rangle &=& |\underbrace{11\ldots 1}_{\frac{T}{3}}\underbrace{00\ldots 0}_{\frac{2T}{3}}\rangle, \\
|C\rangle &=& |\underbrace{00\ldots 0}_{\frac{T}{3}}\underbrace{11\ldots 1}_{\frac{T}{3}}\underbrace{00\ldots 0}_{\frac{T}{3}}\rangle.
\end{eqnarray}
After $\tau_1=\frac{T}{3}$ steps the system is in the state
\begin{equation}
|pre(\tau_1)\rangle = \frac{1}{\sqrt{3}}\left(|x=T/3\rangle\otimes\left(|A\rangle+|C\rangle\right)+|x=-T/3\rangle\otimes |B\rangle\right).
\end{equation}
After $\tau_2=\frac{2T}{3}$ it is in the state 
\begin{equation}
|pre(\tau_2)\rangle = \frac{1}{\sqrt{3}}\left(|x=0\rangle\otimes\left(|B\rangle+|C\rangle\right)+|x=2T/3\rangle\otimes |A\rangle\right).
\end{equation}
Finally, after $T$ steps the system is in the state
\begin{equation}
|pre(T)\rangle = \frac{1}{\sqrt{3}}|x=T/3\rangle\otimes\left(|A\rangle+|B\rangle+|C\rangle\right).
\end{equation}
Next, the system is post-selected in the following state
\begin{equation}
|post\rangle \equiv |post(T)\rangle = \frac{1}{\sqrt{3}}|x=T/3\rangle\otimes\left(|A\rangle+|B\rangle-|C\rangle\right).
\end{equation}
The probability of post selection is $|\langle pre(T)|post(T)\rangle|^2=\frac{1}{9}$. 

The state $|post(T)\rangle$ can be evolved backward in time which allows to make counterfactual predictions about the properties of the system for times $0\leq t \leq T$. In particular, consider what would happen if one measured the position of the particle at times $\tau_1$ and $\tau_2$. The backward evolution implies
\begin{equation}
|post(\tau_1)\rangle = \frac{1}{\sqrt{3}}\left(|x=T/3\rangle\otimes\left(|A\rangle-|C\rangle\right)+|x=-T/3\rangle\otimes |B\rangle\right)
\end{equation}
and 
\begin{equation}
|post(\tau_2)\rangle = \frac{1}{\sqrt{3}}\left(|x=0\rangle\otimes\left(|B\rangle-|C\rangle\right)+|x=2T/3\rangle\otimes |A\rangle\right).
\end{equation}

It is straightforward to evaluate using formula (\ref{ABLprob}) that if at time $\tau_1$ the particle were measured at position  $x=-T/3$ it would definitely be found there. The corresponding conditional probability is $p(\Pi_{x=-T/3}|pre(\tau_1),post(\tau_1)) =1$. On the other hand, at time $\tau_2$ the particle would definitely be found at position $x=2T/3$, since the respective conditional probability is $p(\Pi_{x=2T/3}|pre(\tau_1),post(\tau_1)) =1$.

The above leads to the following paradoxical observation. At time $\tau_1$ the particle was at $x=-T/3$ and at time $\tau_2$ it was at $x=2T/3$. However, the model allows the particle to jump only to neighbouring positions, therefore the distance travelled during the period $\Delta\tau=\tau_2 - \tau_1 = \frac{T}{3}$ could be at most $|\Delta x| \leq \frac{T}{3}$. Nevertheless, the concluded distance travelled by the particle is greater because $\Delta x = T$. Therefore, between $\tau_1$ and $\tau_2$ the average velocity would have to be $v=\frac{\Delta x}{\Delta \tau} = 3$, i.e., the particle would have to jump three positions at a time. 

Note that according to the MCQW model the event corresponding to measuring the particle at $x=-T/3$ at time $\tau_1$ is exclusive to the event corresponding to measuring the particle at $x=2T/3$ at time $\tau_2$. Although the two events happen at different times, one can easily show their exclusivity using the Heisenberg picture
\begin{equation}
U^{\dagger}_{\Delta\tau}\Pi_{x=-T/3}U_{\Delta\tau} \Pi_{x=2T/3}=0,
\end{equation}
where according to (\ref{MCQWevolution})
\begin{equation}
U_{\Delta\tau}=\prod_{i=T/3}^{2T/3 -1} S_i.
\end{equation}
Therefore, the above scenario constitutes a logical PPS paradox which implies contextuality of the MCQW model. 

Unlike in the 3-box scenario, here the system consists of two elements -- position and multi-coin register. It resembles a scenario studied in \cite{SuperluminalQW}, where a superluminal velocity of a particle was inferred from a proper pre- and post-selection. However, here the effect strongly depends on the form of entanglement between the position and the multi-coin register. This is because the intermediate measurements need to be considered at times when paths corresponding to two different multi-coin states meet at the same location. In this case the corresponding PPS states have Schmidt rank two, whereas in general the evolution takes the system to the state with Schmidt rank three. 


\subsection{Scenario 2: Long-distance oscillations}

In the next MCQW scenario we choose
\begin{eqnarray}
|pre\rangle &=& \frac{1}{\sqrt{2s+1}}\left(\sum_{k=1}^{s}|x=2k\rangle\right)\otimes\left(|RL\rangle + |LR\rangle\right) \nonumber \\
&+& \frac{1}{\sqrt{2s+1}}|x=0\rangle\otimes|RL\rangle. \label{even}
\end{eqnarray}
The particle is uniformly distributed over even positions in the region $0 \leq x \leq 2s$ and the multi-coin states $|RL\rangle$ and $|LR\rangle$ are
\begin{equation}
|RL\rangle = |01010101\ldots\rangle,~~~~|LR\rangle = |10101010\ldots\rangle.
\end{equation}
These states lead to oscillations
\begin{eqnarray}
|x\rangle\otimes|RL\rangle \rightarrow |x+1\rangle\otimes|RL\rangle \rightarrow |x\rangle\otimes|RL\rangle \rightarrow \dots \\
|x\rangle\otimes|LR\rangle \rightarrow |x-1\rangle\otimes|LR\rangle \rightarrow |x\rangle\otimes|LR\rangle \rightarrow \dots
\end{eqnarray}
therefore at even time steps the state of the system is the same as (\ref{even}), i.e., $|pre_{even}\rangle \equiv |pre\rangle$, whereas at odd time steps it is
\begin{eqnarray}
|pre_{odd}\rangle &=& \frac{1}{\sqrt{2s+1}}\left(\sum_{k=0}^{s-1}|x=2k+1\rangle\right)\otimes\left(|RL\rangle + |LR\rangle\right) \nonumber \\
&+& \frac{1}{\sqrt{2s+1}}|x=2s+1\rangle\otimes|RL\rangle. \label{odd}
\end{eqnarray}

Next, assume that after an even number of steps $T$ the system is measured and post-selected in the state
\begin{eqnarray}
|post\rangle &=& \frac{1}{\sqrt{2s+1}}\left(\sum_{k=1}^{s}|x=2k\rangle\right)\otimes\left(|RL\rangle - |LR\rangle\right) \nonumber \\
&+& \frac{1}{\sqrt{2s+1}}|x=0\rangle\otimes|RL\rangle. \label{posteven}
\end{eqnarray}
The probability of post-selection is $\langle pre|post\rangle = \frac{1}{s+1}$. As before, the state $|post\rangle$ can be evolved backward in time. One finds that for even time steps $|post_{even}\rangle \equiv |post\rangle$, whereas for odd time steps it yields
\begin{eqnarray}
|post_{odd}\rangle &=& \frac{1}{\sqrt{2s+1}}\left(\sum_{k=0}^{s-1}|x=2k+1\rangle\right)\otimes\left(|RL\rangle - |LR\rangle\right) \nonumber \\
&+& \frac{1}{\sqrt{2s+1}}|x=2s+1\rangle\otimes|RL\rangle. \label{postodd}
\end{eqnarray}

What would be the position of the particle if it were measured at time $\tau$ ($0 \leq \tau \leq T$)? Given position $x$ ($0 < x < 2s+1$) one finds that at arbitrary $\tau$ the pre-selected state implies the multi-coin state of the form $|RL\rangle + |LR\rangle$, whereas the post-selected state implies $|RL\rangle - |LR\rangle$. Since these states are orthogonal the particle cannot be found at these positions. The only possibility is that at even times the particle is at $x=0$, whereas at odd times it is at $x=2s+1$. Note that $p(\Pi_{x=0}|pre_{even},post_{even})=1$ and $p(\Pi_{x=2s+1}|pre_{odd},post_{odd})=1$. This resembles the effect from the previous example. Although the model allows the particle to move only to neighbouring positions, one predicts that it oscillates between $x=0$ and $x=2s+1$ and the period of oscillations is just two steps. As before, these two events are exclusive, therefore the scenario constitutes a logical PPS paradox.

Apart from counterintuitive leaps, the above example exhibits another interesting feature. The pre- and post-selection allows to make definite predictions about the particle's position at all times $\tau=0,1,\ldots, T$. In other words, for any time step $\tau$ there exists at least a single position at which the particle can be found with probability one. Therefore, for each time step one can pick a single position for which the PPS probability is one and create a sequence $\{x_0,x_1,\ldots,x_T\}$. Such a sequence (in the above case $\{0,2s+1, 0,\ldots,2s+1\} $) is a deterministic description of the particle's trajectory and the underlying process can be interpreted as a {\it counterfactual dynamics}. More precisely, although in an actual experiment the measurements are limited only to pre- and post-selection, one can predict that if at time $\tau$ the particle were looked for at position $x_{\tau}$ it would be found there with certainty.


\subsection{Scenario 3: Counterintuitive walks on cycles}

In this scenario we consider a DTQW on a 7-cycle in which the particle always travels either to the left, or to the right (or is in a superposition of both possibilities). Therefore, we do not need to use the MCQW model. We are going to show that at each step one can always find two locations on the cycle, $x_A$ and $x_B$, such that $p(\Pi_{x=x_A}|pre(t),post(t))=p(\Pi_{x=x_B}|pre(t),post(t))=1$. The 7-cycle is the smallest cycle for which we were able to find such an effect. 

\begin{figure}[t]
\center{\includegraphics[scale=0.1,trim={-0cm 0cm 0cm 0cm},clip]{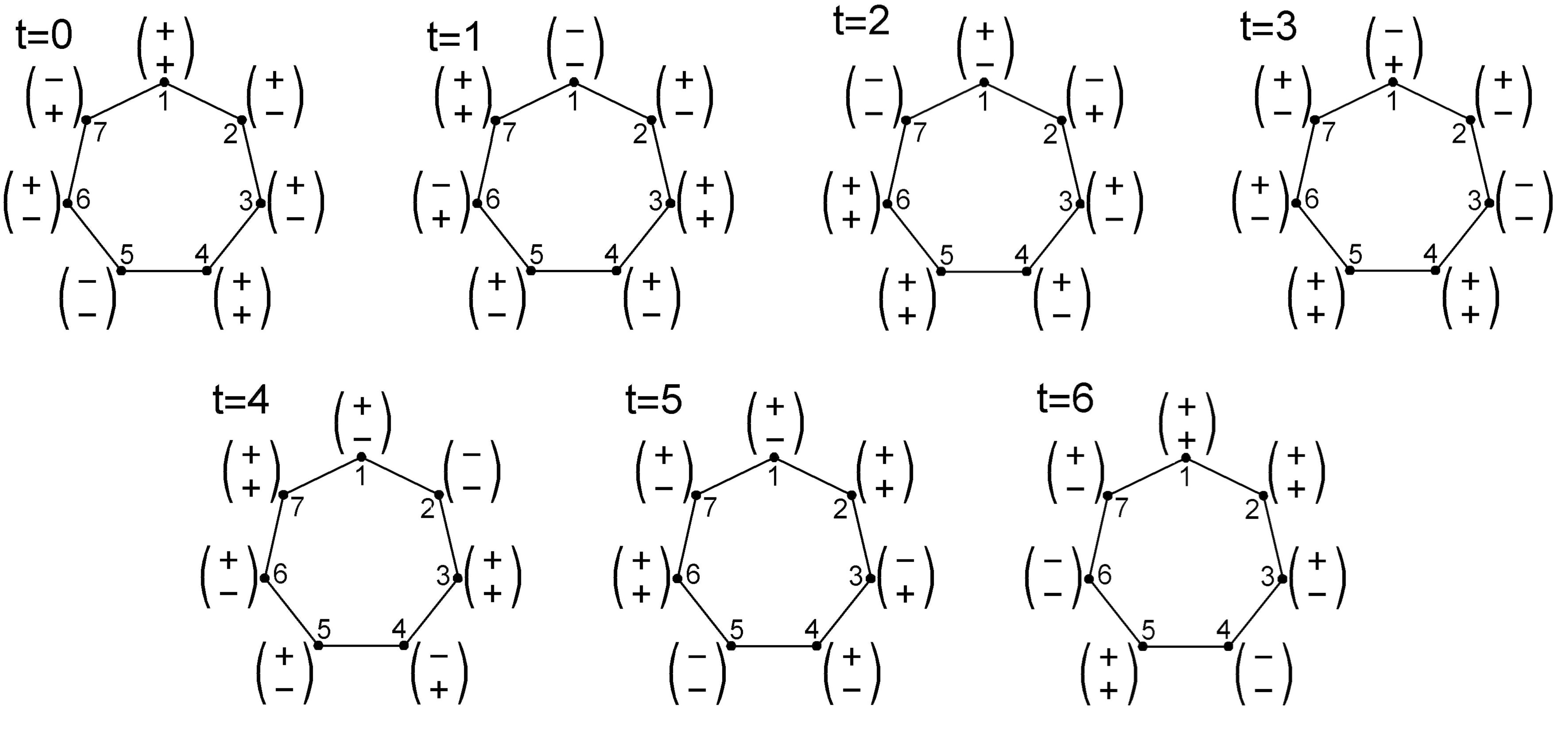}}
\vspace{-0mm}
\caption{Schematic representation of the DTQW evolution on a 7-cycle for the initial state $|pre\rangle$ (\ref{pre7cycle}). For example, for $t=0$ the expresion ${+}\choose{+}$ at position $x=1$ denotes that $|pre(0)\rangle$ contains the term $|x=1\rangle\otimes (|0\rangle + |1\rangle)$, ${-}\choose{+}$ at position $x=7$ denotes that $|pre(0)\rangle$ contains the term $|x=7\rangle\otimes (-|0\rangle + |1\rangle)$, etc. The evolution is periodic and repeats after seven steps. \label{fig3}}
\end{figure}

We set
\begin{eqnarray}
|pre\rangle = |pre(0)\rangle &=& \frac{1}{\sqrt{14}} \left( |1\rangle + |2\rangle + |3\rangle + |4\rangle - |5\rangle + |6\rangle - |7\rangle \right)\otimes |0\rangle \nonumber \\
&+& \frac{1}{\sqrt{14}} \left( |1\rangle - |2\rangle - |3\rangle + |4\rangle - |5\rangle - |6\rangle + |7\rangle \right)\otimes |1\rangle. \label{pre7cycle}
\end{eqnarray}
To shorten the notation we represent the states in the position Hilbert space as $|x=n\rangle \equiv |n\rangle$. The above state represents a particle uniformly spread over all seven positions and over both coin states. The amplitudes have either a phase $+1$ or $-1$, which will be discussed in more details in a moment. Recall, that in the beginning we assumed that the coin toss operation is $C=I_c$, therefore the single step of the DTQW consists only of the conditional translation which transforms (\ref{pre7cycle}) into  
\begin{eqnarray}
|pre(1)\rangle &=& \frac{1}{\sqrt{14}} \left( |2\rangle + |3\rangle + |4\rangle + |5\rangle - |6\rangle + |7\rangle - |1\rangle \right)\otimes |0\rangle \nonumber \\
&+& \frac{1}{\sqrt{14}} \left( |7\rangle - |1\rangle - |2\rangle + |3\rangle - |4\rangle - |5\rangle + |6\rangle \right)\otimes |1\rangle. 
\end{eqnarray}
The next step leads to
\begin{eqnarray}
|pre(2)\rangle &=& \frac{1}{\sqrt{14}} \left( |3\rangle + |4\rangle + |5\rangle + |6\rangle - |7\rangle + |1\rangle - |2\rangle \right)\otimes |0\rangle \nonumber \\
&+& \frac{1}{\sqrt{14}} \left( |6\rangle - |7\rangle - |1\rangle + |2\rangle - |3\rangle - |4\rangle + |5\rangle \right)\otimes |1\rangle 
\end{eqnarray}
and so on. Note that the evolution is periodic, i.e., $|pre(7)\rangle = |pre(0)\rangle$ and in general $|pre(t)\rangle = |pre(t+7k)\rangle$, where $k$ is an arbitrary integer. All seven steps are schematically represented in Fig. \ref{fig3}.

Next, assume that after an arbitrary number of steps $T$ the system is post-selected in the state
\begin{eqnarray}
|post\rangle = \frac{1}{\sqrt{14}} \left( |1\rangle + |2\rangle + |3\rangle + |4\rangle + |5\rangle + |6\rangle + |7\rangle \right)\otimes (|0\rangle + |1\rangle), \label{post7cycle}
\end{eqnarray}
which is an eigenstate of the DTQW single step operator, therefore its backward evolution is trivial $|post(t)\rangle = |post\rangle$. This choice allows us to radically simplify our search for PPS paradoxes. In addition, note that $|\langle pre(t)|post\rangle|^2 = \frac{1}{49}$ for all $t$.

In order to observe logical PPS paradoxes in the above system let us note that the post-selected state is an equal superposition of 14 terms, each having the same phase $+1$. On the other hand, the pre-selected state, although evolving, is always an equal superposition consisting of eight terms with phase $+1$ and six terms with phase $-1$. If the particle were found at some position $x$, the post-selected state would collapse onto 
\begin{equation}\label{collapse7cycle}
\frac{1}{\sqrt{2}}|x\rangle\otimes(|0\rangle+|1\rangle)
\end{equation}
and the pre-selected state would collapse onto 
\begin{equation}
\frac{1}{\sqrt{2}}|x\rangle\otimes(\alpha(x,t)|0\rangle+\beta(x,t)|1\rangle),
\end{equation}
where $\alpha(x,t),\beta(x,t)=\pm 1$ are position- and time-dependent phase factors. If the particle were looked for at position $x$, but were not found there, the corresponding states would collapse onto
\begin{equation}
|post_{\bar{x}}\rangle = \frac{1}{\sqrt{12}}\sum_{x'\neq x} |x'\rangle \otimes (|0\rangle + |1\rangle)
\end{equation}
and
\begin{equation}
|pre(t)_{\bar{x}}\rangle = \frac{1}{\sqrt{12}}\sum_{x'\neq x} |x'\rangle \otimes (\alpha(x',t)|0\rangle + \beta(x',t)|1\rangle).
\end{equation}
The crucial observation is that in order to get $p(\Pi_{x}|pre(t),post)=1$ the following must hold: 
\begin{equation}
0=\langle pre(t)_{\bar{x}}|post_{\bar{x}}\rangle = \sum_{x'\neq x}  (\alpha(x',t) + \beta(x',t)).
\end{equation}
This can only happen if exactly six of the amplitudes  $\alpha(x',t)$ and $\beta(x',t)$ are $+1$ and the remaining six are $-1$. However, this implies that if the particle were found at $x$ the pre-selected state and post-selected state would collapse onto the same state (\ref{collapse7cycle}). 

In simple words, the following rule applies: {\it in order to get $p(\Pi_{x}|pre(t),post)=1$ the phases of the amplitudes corresponding to $|x\rangle\otimes|0\rangle$ and $|x\rangle\otimes|1\rangle$ in the state $|pre(t)\rangle$ need to be both $+1$}. Using this rule one can easily find logical PPS paradoxes using the schematic representation of the evolution in Fig. \ref{fig3}. In this figure the positions for which the particle would be found with certainty are marked with ${+}\choose{+}$. The PPS paradox requires that $p(\Pi_{x}|pre(t),post)=1$ for at least two positions. One can easily find that if at time $t$ one were given a choice to look for the particle at a single position one would find it with certainty at
\begin{itemize}
\item 1 and 4 for $t=7k$,
\item 3 and 7 for $t=7k+1$,
\item 5 and 6 for $t=7k+2$,
\item 4 and 5 for $t=7k+3$,
\item 3 and 7 for $t=7k+4$,
\item 2 and 6 for $t=7k+5$,
\item 1, 2 and 5 for $t=7k+6$.
\end{itemize} 

The paradoxical observation is that at all times the particle seems to be in more than one position simultaneously. Moreover, just like in the previous example, one can also choose a sequence of positions $\{x_0,x_1,\ldots,x_T\}$ describing a counterfactual dynamics. However, unlike in the previous case, this time there is more than one such sequence, because for each time step one can pick one of two, or three, positions. For example, for $T=7$ there are $3 \times 2^6 = 192$ different sequences. What is important, none of them describes a particle hopping one step to the right, or left, since there is always at least one counterintuitive leap, like $\{1,7,6,5,3,2,1\}$. Finally, even though the actual evolution is periodic, the counterfactual dynamics does not need to be so. For example, for $T=14$ one can choose the following sequence $\{1,7,6,5,3,2,1,1,3,5,4,7,6,2\}$.  


\subsection{Scenario 4: Walks on disconnected graphs}

In the final example we consider a DTQW on three disconnected 3-cycles, call them A, B and C. As in the previous example, the coin toss operator is $I_c$, therefore the one step evolution operator consists solely of a conditional shift 
\begin{equation}\label{evolutiondisconnected}
U=\sum_{X=A,B,C}\sum_{x=1}^3(|x_X+1\rangle\langle x_X|\otimes|0\rangle\langle 0|+|x_X-1\rangle\langle x_X|\otimes|1\rangle\langle 1|),
\end{equation}
where $x_X$ denotes the position $x$ on the 3-cycle $X$ (e.g. $1_A$ means position $x=1$ on the 3-cycle A) and we assume periodic boundary conditions $x_X \equiv x_X~mod~3$. The above dynamics does not allow to hop from A to B, etc. However, we are going to show a PPS paradox for which the particle can be at two or three different cycles at the same time and can hop between them. This resembles the paradox of a disappearing and reappearing particle \cite{DRP}.  

\begin{figure}[t]
\center{\includegraphics[scale=0.1,trim={-0cm 0cm 0cm 0cm},clip]{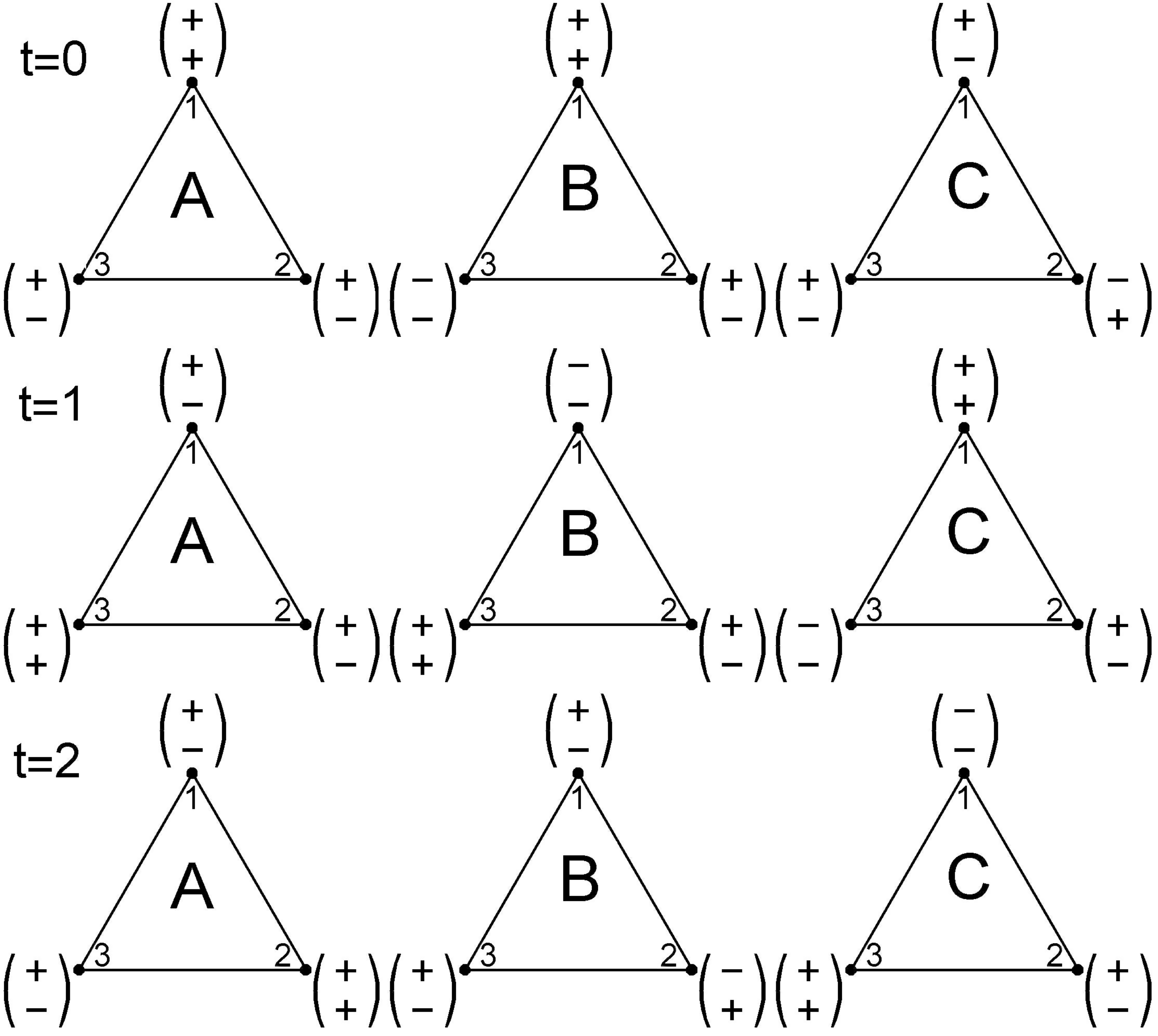}}
\vspace{-0mm}
\caption{The schematic representation of the DCQW evolution on three 3-cycles for the pre-selected state (\ref{predisconnected}). The notation is the same as in Fig. \ref{fig3}. \label{fig4}}
\end{figure}

Let us pre-select the state of the system in a uniform superposition
\begin{eqnarray}
|pre\rangle \equiv |pre(0)\rangle = \label{predisconnected} \\
\frac{1}{\sqrt{18}}(|1_A\rangle + |2_A\rangle + |3_A\rangle+|1_B\rangle + |2_B\rangle - |3_B\rangle+|1_C\rangle - |2_C\rangle + |3_C\rangle)\otimes|0\rangle + \nonumber \\ 
\frac{1}{\sqrt{18}}(|1_A\rangle - |2_A\rangle - |3_A\rangle+|1_B\rangle - |2_B\rangle - |3_B\rangle-|1_C\rangle + |2_C\rangle - |3_C\rangle)\otimes|1\rangle. \nonumber
\end{eqnarray}
It evolves into
\begin{eqnarray}
|pre(1)\rangle = \\
\frac{1}{\sqrt{18}}(|2_A\rangle + |3_A\rangle + |1_A\rangle+|2_B\rangle + |3_B\rangle - |1_B\rangle+|2_C\rangle - |3_C\rangle + |1_C\rangle)\otimes|0\rangle + \nonumber \\ 
\frac{1}{\sqrt{18}}(|3_A\rangle - |1_A\rangle - |2_A\rangle+|3_B\rangle - |1_B\rangle - |2_B\rangle-|3_C\rangle + |1_C\rangle - |2_C\rangle)\otimes|1\rangle, \nonumber
\end{eqnarray}
and then into
\begin{eqnarray}
|pre(2)\rangle = \\
\frac{1}{\sqrt{18}}(|3_A\rangle + |1_A\rangle + |2_A\rangle+|3_B\rangle + |1_B\rangle - |2_B\rangle+|3_C\rangle - |1_C\rangle + |2_C\rangle)\otimes|0\rangle + \nonumber \\ 
\frac{1}{\sqrt{18}}(|2_A\rangle - |3_A\rangle - |1_A\rangle + |2_B\rangle - |3_B\rangle - |1_B\rangle-|2_C\rangle + |3_C\rangle - |1_C\rangle)\otimes|1\rangle. \nonumber
\end{eqnarray}
The evolution is periodic and the period is three $|pre(t+3k)\rangle \equiv |pre(t)\rangle$. It is schematically represented in Fig. \ref{fig4}, where we used the same notation as in the previous example.

After $T$ steps let us post-select the system in the state
\begin{eqnarray}
|post\rangle  &=& \frac{1}{\sqrt{18}}\left(|1_A\rangle + |2_A\rangle + |3_A\rangle + |1_B\rangle + |2_B\rangle + |3_B\rangle \right. \nonumber \\
& & \left. +|1_C\rangle + |2_C\rangle + |3_C\rangle \right) \otimes  (|0\rangle + |1\rangle).  
\end{eqnarray}
Just like in the previous example, the post-selected state is the eigenstate of the evolution and $|\langle pre(t)|post\rangle|^2 =\frac{1}{81}$ for all $t$. The mechanism behind the PPS paradox is also the same as before. The post-selected state is an even superposition with 18 terms, each having phase $+1$, whereas the pre-selected state, although evolving, is always an even superposition with ten terms having phase $+1$ and eight terms having phase $-1$. Therefore, measurement of the particle at a position $x$, for which both coin states have phases $+1$, i.e., $|x\rangle\otimes(|0\rangle+|1\rangle)$, will always have a conditional probability $p(\Pi_x|pre(t),post)=1$. In Fig. \ref{fig4} such positions are marked by ${+}\choose{+}$. With the help of this graphical representation it is easy to see that if at time $t$ one were given a choice to look for a particle at a single position one would find it with certainty at
\begin{itemize}
\item $1_A$ and $1_B$ for $t=3k$,
\item $3_A$, $3_B$ and $3_C$ for $t=3k+1$, 
\item $2_A$ and $3_C$ for $t=3k+2$.
\end{itemize}
The above gives rise to a logical PPS paradox since the particle could be found with certainty at more than one position. In addition, in the counterfactual dynamics the particle can hop between the 3-cycles, which is not allowed in the actual dynamics. Apart from intuitive periodic counterfactual dynamical sequences of positions, like $\{1_A,3_A,2_A,1_A,3_A,2_A,\ldots\}$, one can find counterintuitive aperiodic ones, like $\{1_B,3_C,2_A,1_A,3_B,3_C,\ldots\}$.

Finally, let us note that although the actual dynamics of the system does not allow to hop between the 3-cycles both pre-selection and post-selection require a joint measurement on all 3-cycles. If preparation and final measurement were limited to a single cycle only, one would never be able to observe the above paradoxical situation. 


\section{Conclusions}

We showed that logical PPS paradoxes can be observed in quantum walks, which implies contextuality of the model \cite{Pusey,PuseyLeifer}. This supports the result in \cite{QWLG} that single-partite quantum walk dynamics is non-classical, however we clearly pointed out that such a claim requires specific assumptions (no hidden memory in the system). The discrete-time evolution allows to make paradoxical predictions at every time step and for natural position measurements, which makes quantum walks well suited models to study such effects. The reason is that logical PPS paradoxes require that $|pre(t)\rangle$ and $|post(t)\rangle$ are of a special form. Discrete dynamics allows to jump from one pair of such states to another in one step, whereas continuous-time dynamics transforms continuously one pair into the other. Therefore, if one is restricted solely to compatible discrete-outcome measurements, the logical paradox cannot be observed while the transformation takes place. 

In addition, we were able to define time sequences of position measurements that would certainly detect the particle, if an intermediate measurement was performed. We suggested that such sequences can be interpreted as counterfactual dynamics. Apart from counterfactual description, the question remains whether such dynamics can provide any additional insight into the system's behaviour.

Finally, although for a simplicity of the presentation we chose to consider a special type of quantum walks in which the coin degree of freedom does not evolve, we note that one can also construct PPS logical paradoxes in standard quantum walks. A simple example is a Hadamard walk on 3-cycle. If at $t=0$ the system was prepared in the state 
\begin{equation}
|pre(0)\rangle = \frac{1}{\sqrt{6}}(|1\rangle + |2\rangle + |1\rangle)\otimes(|0\rangle + |1\rangle)
\end{equation}
and at $t=3$ it was measured in the state
\begin{equation}
|post(3)\rangle = \frac{1}{\sqrt{3}}|3\rangle\otimes(|0\rangle + |1\rangle)-\frac{1}{\sqrt{3}}|2\rangle\otimes |1\rangle,
\end{equation}
then if at time $0 \leq t \leq 3$ one were given a choice to look for a particle at a single position one would find it with certainty at
\begin{itemize}
\item 2 and 3 for $t=0$,
\item 1 and 3 for $t=1$,
\item 2 for $t=2$,
\item 3 for $t=3$.
\end{itemize} 


\section*{Acknowledgements}

We would like to thank M. Karczewski for discussions and helpful comments. This work is supported by the Ministry of Science and Higher Education in Poland (science funding scheme 2016-2017 project no. 0415/IP3/2016/74).



\end{document}